\documentclass[aps,prb,twocolumn,showpacs]{revtex4}
\usepackage{epsfig}
\usepackage{amsmath}
\usepackage{amssymb}
\usepackage{color}  % optional for remarks & changes
\usepackage{ulem}   % optional for remarks & changes
\begin{document}
\title{Quantum lattice dynamical effects on the single-particle excitations\\
in 1D Mott and Peierls insulators}
\author{H.~Fehske,$^{1}$ G. Wellein,$^{2}$ G. Hager,$^{2}$
  A. Wei{\ss}e,$^{3}$
and A. R. Bishop$^{4}$}
\affiliation{$^{1}$Institut f\"ur Physik, Ernst-Moritz-Arndt 
Universit\"at Greifswald, D-17487 Greifswald, Germany}
\affiliation{$^{2}$Regionales Rechenzentrum Erlangen, 
Universit\"at Erlangen, D-91058 Erlangen, Germany}
\affiliation{$^{3}$School of Physics, The University of New South Wales,
Sydney, NSW 2052, Australia}
\affiliation{\mbox{$^{4}$Theoretical 
Division and Center for Nonlinear Studies, 
  Los Alamos National Laboratory, Los Alamos, New Mexico 87545}}
\date{\today} 
\begin{abstract}
As a generic model describing quasi-one-dimensional Mott 
and Peierls insulators, we investigate the Holstein-Hubbard model 
for half-filled bands using numerical techniques. Combining 
Lanczos diagonalization with Chebyshev moment 
expansion we calculate exactly the photoemission and inverse photoemission 
spectra and use these to establish the phase diagram of the model. 
While polaronic features emerge only at strong electron-phonon 
couplings, pronounced phonon signatures, such as multi-quanta band
states, can be found in the Mott insulating regime as well. 
In order to corroborate the Mott to Peierls transition scenario, 
we determine the spin and charge excitation gaps by a 
finite-size scaling analysis based on density-matrix 
renormalization group calculations.    
\end{abstract}
\maketitle
The one-dimensional (1D) Holstein-Hubbard model (HHM) has been used 
extensively to describe for novel low-dimensional materials, e.g., conjugated 
polymers, organic charge transfer salts or halogen-bridged transition metal 
complexes\cite{TNYS90}, and the associated metal-insulator\cite{MP,CSCDG03} 
and insulator-insulator transitions\cite{FWWGBB02,SSC03}.\cite{BVL95} 
The HHM accounts for a tight-binding
electron band, intra-site Coulomb repulsion between 
electrons of opposite spin, and a local coupling of the 
charge carriers to the phonon system: 
\begin{eqnarray}
\label{hhm}
H&=&- t\sum_{i,\sigma}(c^{\dagger}_{i\sigma}c^{}_{i+1 \sigma}
+\mbox{H.c.})+U\sum_i n_{i\uparrow}n_{i\downarrow}\nonumber\\
&&-\sqrt{\varepsilon_p\omega_0}\sum_{i,\sigma}(b_i^{\dagger}
+b_i^{})n_{i\sigma}+\omega_0
\sum_{i} b_i^{\dagger} b_i^{}\,. 
\end{eqnarray}
Here, $c^{\dagger}_{i\sigma}$ ($c^{}_{i\sigma}$) denote
fermionic creation (annihilation) operators of spin-$\sigma$ electrons
($\sigma=\uparrow,\,\downarrow$) on a 1D lattice with $N$ sites, 
$n_{i\sigma}=c^{\dagger}_{i\sigma}c^{}_{i\sigma}$,
and $b^{\dagger}_{i}$ ($b^{}_{i}$) are the corresponding bosonic 
operators for a dispersionsless optical phonon
with frequency $\omega_0$.

The physics of the model~(\ref{hhm}) is governed by the competition between
electron itinerancy ($\propto W=4t$) on the one hand and 
electron-electron ($\propto u=U/4t$) and electron-phonon 
($\propto \lambda=\varepsilon_p/2t$;   
$\varepsilon_p$ is the polaron shift) 
interactions on the other hand, which both tend to immobilize 
the charge carriers. At least for the half-filled band case
($\sum_{i,\sigma}n_{i\sigma}=N_{el}=N$),  
Mott insulator (MI) or Peierls insulator (PI) states are 
expected to be favored over the metallic state at temperature 
$T=0$ (see Fig.~1). The correlated MI shows pronounced spin-density-wave (SDW)
fluctuations but has continuous symmetry. It therefore
exhibits no long-range order in 1D. In contrast the PI is 
characterized by dominant charge-density-wave (CDW) correlations 
and true long-range order because a discrete symmetry is broken.      
While the gaps to both spin ($\Delta_s$) and charge ($\Delta_c$) excitations
are finite in the PI, the spin gap vanishes in the 1D MI,
which is related to spin charge separation.
\begin{figure}[t]
  \includegraphics[width=.85\linewidth,clip]{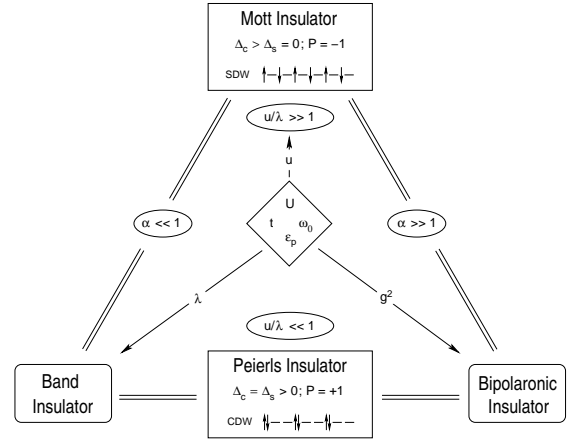}
  \caption{Schematic phase diagram of the Holstein-Hubbard model
    at half-filling.}\label{f1}\vspace*{-0.3cm}
\end{figure}  
In a strict sense these results hold in the adiabatic limit 
($\omega_0=0$) for ``$U$-only'' (Hubbard model) and 
``$\lambda$-only'' (Peierls model)  
parameters. 
At finite phonon frequency and $U=0$ (Holstein model) 
a critical electron-phonon (EP) coupling is required to set 
up the CDW phase characterized by alternating doubly occupied and empty 
sites.\cite{BMH98} The concomitant gap formation and metal-insulator 
transition has recently been studied in the limit of infinite 
dimensions.\cite{MHB02}  
Depending on the adiabaticity ratio 
$\alpha=\omega_0/t$ the PI represents a traditional 
band insulator ($\alpha\ll 1$) or  a bipolaronic 
insulator ($\alpha\gg 1$, $g^2=\varepsilon_p/\omega_0\gg1$).\cite{remark1}   
Although for the more general HHM the situation is much less clear, 
we expect that the features of the insulating phase will  
depend markedly on the ratio of Coulomb and EP interactions 
$u/\lambda$, allowing for quantum phase transitions between 
insulating phases. Indeed, based on recent numerical results 
for the staggered spin- and charge-structure factors, it has 
been argued that the HHM shows a crossover between Mott and 
Peierls insulating phases at
$u/\lambda\simeq 1$.\cite{FKSW03} 
More precisely, for finite 
periodic chains, the MI-PI quantum phase transition 
could be identified by a ground-state level crossing associated 
with a change in the parity eigenvalue $P$.\cite{FWWGBB02}  
Note that this scenario differs from the (weakly
interacting) HHM with frozen phonons~\cite{FKSW03}, where there is
strong evidence in favor of two quantum critical points, as in the
ionic Hubbard model.\cite{BJKS03,IHM}

The aim of this work is to establish the physical 
picture developed to date for the interplay of spin, 
charge and lattice degrees of freedom in the
1D Holstein Hubbard model. In particular we attempt to verify
the proposed phase diagram by examining the single-particle excitations. 
For these purposes we employ Lanczos exact diagonalization (ED)\cite{ED}, 
kernel polynomial\cite{KPM}, 
and density-matrix renormalization group (DMRG)\cite{White,Hager} methods 
to determine the (inverse) photoemission spectra 
as well as the spin and charge excitation gaps. 
These quasi-exact numerical techniques allow  
us to obtain reliable results for all interaction 
strengths with the full quantum dynamics of phonons 
taken into account. Exact diagonalizations are seriously limited in
  achievable system sizes $N$, but have the advantage that spectral
  quantities are easily accessible. Complementary, the DMRG algorithm
  yields specific eigenstates of large systems by implementing a
  renormalization scheme and an optimal truncation of the Hilbert 
  space. Thus it permits for a finite-size analysis of the
  ground-state energies in different particle and spin sectors, which
  is required to determine the behavior of the various excitation
  gaps in the thermodynamic limit. 

We begin  by studying the spectral density 
of single-particle excitations associated with 
the injection of a spin-$\sigma$ electron with wave number 
$K$, $A_{K\sigma}^{+}(\omega)$ (inverse photoemission (IPE)),
and the corresponding quantity 
for the emission of an electron, $A_{K\sigma}^{-}(\omega)$ 
(photoemission (PE)), where  
\begin{eqnarray}
A_{K\sigma}^{\pm}(\omega)&=& \sum_{m} 
|\langle \psi_m^{(N_{el}\pm1)}|c_{K\sigma}^{\pm} 
|\psi_0^{(N_{el})}\rangle|^2\nonumber \\
&&\qquad\times\delta [\,\omega\mp(E_m^{(N_{el}\pm1)}-E_0^{(N_{el})})] 
\label{ipe}
\end{eqnarray}
with $c_{K\sigma}^{+}=c_{K\sigma}^{\dagger}$  
and $c_{K\sigma}^{-}=c_{K\sigma}^{}$. 
$|\psi_0^{(N_{el})}\rangle$ is the ground state of the system
with $N_{el}$ electrons and $|\psi_m^{(N_{el}\pm 1)}\rangle$ are 
eigenstates of the $(N_{el}\pm 1)$-particle system. 
$E_0^{(N_{el})}$ and $E_m^{(N_{el}\pm1)}$ are the corresponding
energies. Adding the spectral densities of (photo-) emission and absorption 
we obtain the spectral function 
$A_{K\sigma}(\omega)=A_{K\sigma}^+(\omega)+ A_{K\sigma}^-(\omega)$,
which obeys various sum rules and allows for a connection to
angle-resolved photoemission spectroscopy (ARPES). The simplest sum
rule, $\int_{-\infty}^{\infty} A_{K\sigma}(\omega) d\omega=1$,
reflects the normalization of $A_{K\sigma}(\omega)$ but is not
useful for ARPES since it involves both occupied and unoccupied states.
$\sum_\sigma\int_{-\infty}^{\infty}
n_F(\omega) A_{K\sigma} (\omega) d\omega = n(K)$ (where
$n_F(\omega)$ is the Fermi function) is more important, since it 
relates the ARPES intensity to the number of electrons 
in a momentum state~$K$: $n(K)=  
\sum_\sigma \langle c_{K\sigma}^\dagger c_{K\sigma}^{}\rangle $. 
The ED results presented for $A_{K\sigma}^{\pm}(\omega)$ 
in the following were obtained for an eight-site 
system with periodic boundary conditions.\cite{remark2}  

Let us first consider the MI regime. Figure~2 displays 
the IPE and PE spectra for the HHM at the allowed
wave numbers of our finite system:
$K=0,\, \pm\pi/4,\,\pm\pi/2, \pm 3\pi/4$, 
and $\pi$. To reliably monitor  
a possible band splitting induced by the Hubbard and
EP couplings at half-filling it is necessary to guarantee that 
the Fermi momenta $K_F=\pm\pi/2$ are occupied, which is  
the case for $N=4l$ ($l$ integer, periodic boundary conditions).
The most prominent feature we observe in the MI regime is 
the opening of a gap at $K=\pm \pi/2$, indicating massive charge 
excitations. A comparison with the results obtained for 
the pure Hubbard model classifies this 
gap as the Mott-Hubbard correlation gap. 
Its value $\Delta /t \simeq 3.25$ almost coincides with
the optical gap $\Delta_{opt}$ we determined by  
evaluating the regular part of the optical conductivity
for the same parameters. The dispersion of the lower (upper) Hubbard band
can be derived tracing the uppermost (lowest) excitations 
in each $K$ sector. 
%% Of course, the degeneracy of 
%% wave numbers 0 and $\pi$ at $u\gg 1$, being  
%% related to the doubling of the
%% unit cell in the MI phase, will become complete in 
%% the $N\to\infty$-limit only. 
Due to the finiteness 
of our system and the rather moderate value $u=1.5$,
PE (IPE) excitations with $K=\pm 3\pi/4$ and $\pi$
($K=\pm \pi/4$ and $0$) have still finite spectral weight.
This can be seen from the integrated spectral densities 
$S_{K\sigma}^{\pm}(\omega)=\int_{\mp \infty}^{\omega}d\omega^\prime 
A_{K\sigma}^{\pm }(\omega^\prime)$, which, in addition to the sum rule
$S_{K\sigma}^{-}(-\infty)+S_{K\sigma}^{+}(\infty)=1$,
satisfy the relations 
 $S^{\pm}_{K,\sigma} (\pm \infty ) + S^{\pm}_{\pi-K,\sigma} (\pm \infty ) = 1$ 
($K\geq 0$).  
Since the spectral weight of the PE excitations with $K>\pi/2$
is expected to vanish as $N$ goes to infinity for $u\gg 1$, 
the lower Hubbard band will be completely filled 
($\sum_{|K|\leq K_F,\sigma}\int_{\infty}^{-\infty}d\omega 
A_{K\sigma}^{-}(\omega)\simeq N_{el}$), and consequently
the system behaves as an insulator at $T=0$.\cite{Pea94}   
As a result of the coupling to the phonon system the electronic 
levels in each $K$ sector split,  creating phonon 
side bands. The distinct peaks are separated by
multiples of the bare phonon frequency and 
can be assigned to relaxation processes of the $Q=0$ phonon
modes.\cite{Ro97} 
%transitions with different 
%numbers of (zero-momentum) phonons excited.\cite{Ro97} 
The number of phonons involved is controlled by $g^2$.  
$S_{K\sigma}^{\pm}(\omega)$ shows clearly that 
the total spectral weight of the resulting excitation
bands equals the weight of the respective electronic 
excitations in the pure Hubbard model.
Interestingly, mediated by $Q\neq 0$ phonons,
there appear ``shadows'' of the band belonging to a 
dominant electronic excitation in a certain $K$ sector  
in other $K$ sectors, giving rise to a weak   
``breather-like'' excitation\cite{WBGS98}, which is almost   
dispersionsless in the Brillouin zone. 

If we decrease the Hubbard interaction at fixed EP coupling 
strength the Mott-Hubbard gap weakens and  finally  closes at about 
$(u/\lambda)_c\simeq 1$, which marks the MI-PI crossover. 
This is the situation shown in  Fig.~3. 
Approaching the critical point from above and 
below, the ground state and the first excited state become degenerate.
These states have different eigenvalues $P$ of the 
site-inversion operator $Pc_{i\sigma}^\dagger P^\dagger
= c_{N-i\,\sigma}^\dagger$ ($i=0,\ldots,N-1$) 
and we have verified that the ground-state site parity is 
$P=+1$ in the MI and $P=-1$ in the PI. Obviously   
the critical point is characterized by gapless charge
excitations at the Fermi momenta  
but should not be considered
\newpage 
\begin{figure}[t]
\label{f2}
\epsfig{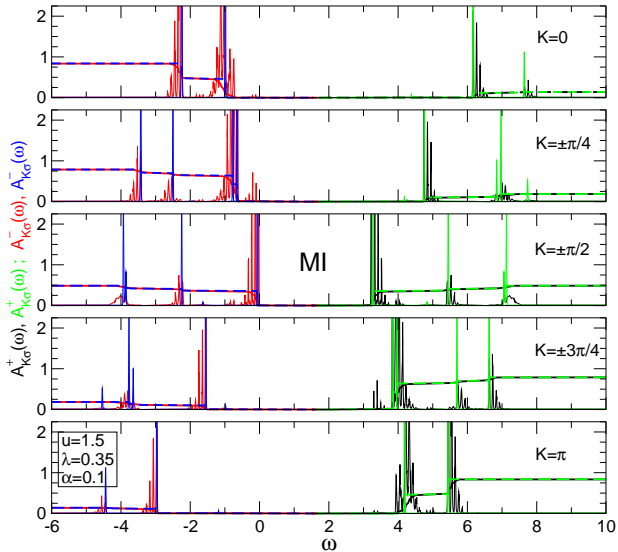}
\caption{Wave-number-resolved spectral densities  
for photoemission ($A_{K\sigma}^-(\omega)$; red lines) 
and inverse photoemission ($A_{K\sigma}^+(\omega)$; black lines)
in the Mott insulating state ($u/\lambda\gg 1$).  
The corresponding integrated densities
$S_{K\sigma}^{\pm}(\omega)$ are given by dashed lines. 
Data for the pure Hubbard model (blue and green lines)
were shifted by $-(\varepsilon_p N^2_{el}/N)$ and included  
for comparison.} 
\end{figure}
\begin{figure}[ht]
\label{f3}
\epsfig{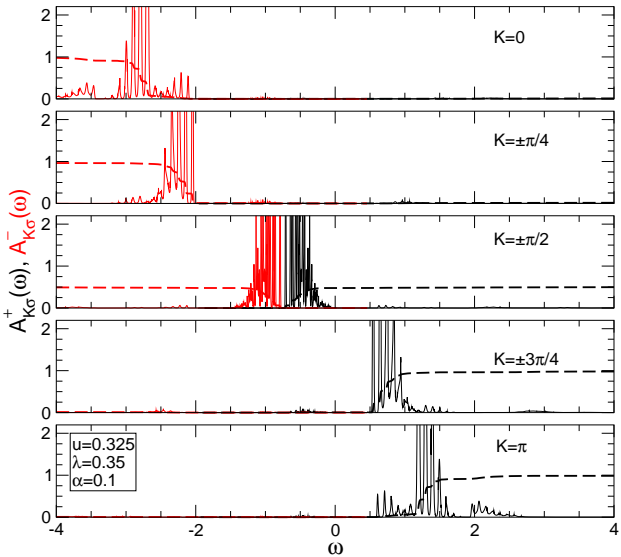}
\caption{PE (red lines) and IPE (red lines) spectra
near the Mott insulator Peierls insulator transition  point 
($u\simeq \lambda$).}
\end{figure} 
\begin{figure}[t!]
\label{f4}
\epsfig{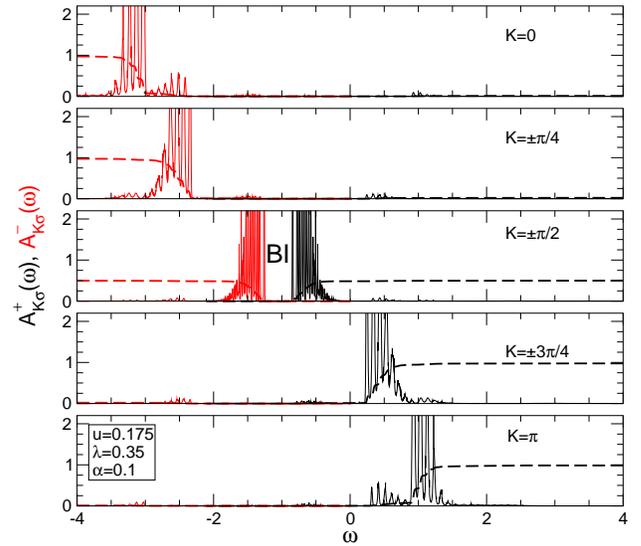}\\[0.7cm]
\epsfig{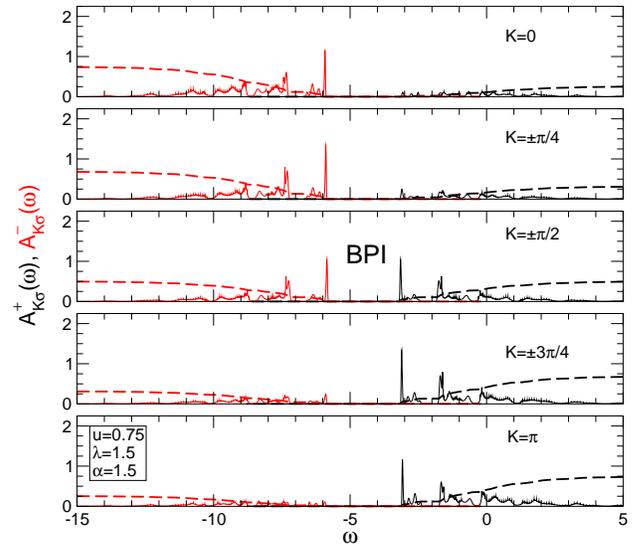}
 \caption{PE and IPE spectra in the Peierls phase ($u\ll \lambda$).
The upper (lower) panels show typical results obtained  
for the case of a band insulator (BI) at $\alpha\ll 1$ 
and bipolaronic insulator (BPI) at $\alpha\gg 1$, respectively.}
\end{figure}
\noindent
as metallic because the Drude weight is  ill-defined.~\cite{BJKS03,Ko64} 

If the Hubbard interaction is further reduced, i.e. 
the EP coupling overcomes the on-site Coulomb repulsion,
a CDW accompanied by a dimerization of the lattice develops.
As a result the electronic band structure becomes gapped again 
(see Fig.~4 (upper panel)). 
The form of the spectra, however, is quite different from MI case. 
While in the MI regime the lowest peak in each $K$ sector 
is clearly the dominant one, in the BI phase rather broad (I)PE 
signatures appear. Within these excitation bands the spectral 
weight is almost uniformly distributed, 
which is a clear signature of multi-phonon
absorption and emission processes that accompany every
single-particle excitations in the PI. 
The lineshape then reflects the (Poisson-like) distribution 
of the phonons in the ground state. 
Again low-intensity ``shadow bands'' become visible. 
Remarkably, now the cumulative spectral weight 
$\sum_{|K|\leq\pi/2} S^-_{K\sigma}(-\infty)$ 
gives nearly the total number $N_{el}$ of electrons, i.e., 
the finite-size effect mentioned above for the MI 
are much less pronounced in the BI state. 
The situation changes radically if the insulating behavior
is associated with localized bipolarons forming a CDW state 
(see Fig.~4, lower panel). Due to strong polaronic effects
an almost flat band dispersion results with exponentially  
small (electronic) quasiparticle weight. Now the dominant 
peaks in the incoherent part of the (I)PE spectra
are related to multiples of the (large) bare phonon
frequency broadened by electronic excitations.      
\begin{figure}[t]
    \epsfig{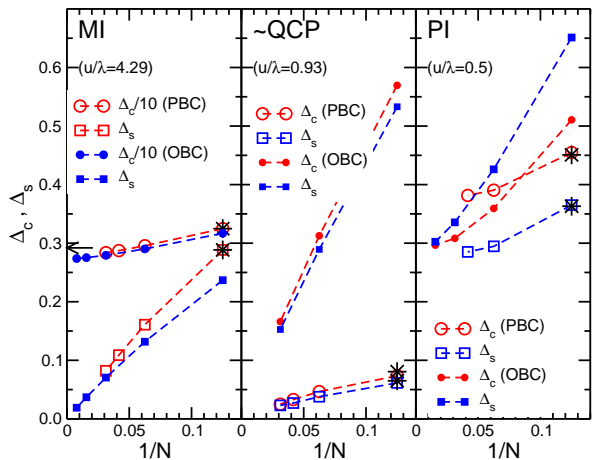}
   \caption{DMRG finite-size scaling of spin- and charge excitation
     gaps in the HHM with dynamical phonons ($\lambda=0.35,\;\alpha=0.1$). 
Note the different scale of $\Delta_{c}$ in the MI phase.
Open and filled symbols denote DMRG results for   
periodic (PBC) and open (OPC) boundary conditions,
respectively. The accessible system sizes are smaller at
larger $\lambda/u$, where an increasing number 
of (phononic) pseudosites is required to reach  
convergence with respect to the phonons.  
Stars represent the ED results for the eight-site system.
The arrow marks the value of the optical gap $\Delta_{opt}$
for the Bethe-ansatz solvable 1D Hubbard model, 
which is given  by $\Delta_{opt}/4t = u - 1 + \ln(2)/2u$
in the limit of large $u>1$.\cite{JGE00}} 
\end{figure}

Since many-body gaps to excited states 
form the basis for making contact with experimentally measurable 
excitation gaps and can also be used to characterize
different phases of the HHM, we finally determine the  
charge and spin gaps,
%$\Delta_c=E_0^{(N+1)}(\tfrac{1}{2})+E_0^{(N-1)}(-\tfrac{1}{2})-2E_0^{(N)}(0)$ 
%and spin gap $\Delta_s=E_0^{(N)}(1)-E_0^{(N)}(0)$  
 \begin{eqnarray}
 \label{cg}
 \Delta_c&=&E_0^{(N+1)}
(\tfrac{1}{2})+E_0^{(N-1)}(-\tfrac{1}{2})-2E_0^{(N)}(0)\\
 \label{sg}
 \Delta_s&=&E_0^{(N)}(1)-E_0^{(N)}(0)\,,
 \end{eqnarray}
using DMRG.\cite{remark3} 
Here $E_0^{(M)}(S^z)$ is the ground-state 
energy of the HHM with M partcles in the sector with  
total spin-$z$ component $S^z$.

Obviously the finite-size scaling presented in Fig.~5 
for $\Delta_{c/s}$ substantiates our introductory discussion of the
phase diagram (cf. Fig.~1). $\Delta_c$ and $\Delta_s$ are finite
in the PI and will converge further as $N\to \infty$. 
Both gaps seem to vanish at the quantum phase transition point
of the HHM with finite-frequency phonons, but in the critical
region the finite-size scaling is extremely delicate.
In the MI we found a finite charge excitation gap, which
in the limit $u/\lambda\gg 1$ scales to the optical gap 
of the Hubbard model, whereas the extrapolated spin gap 
remains zero.\cite{remark4}   

In summary, we have presented a comprehensive picture of
the physical properties of the 1D half-filled 
finite-phonon frequency Holstein-Hubbard model.
With respect to the metal the electron-electron coupling
favors the Mott insulating state whereas the electron-phonon
interaction is responsible for the Peierls insulator to occur.
The PI typifies a band insulator in the adiabatic 
weak-to-intermediate coupling range or a bipolaronic
insulator for non-to-antiadiabatic strong-coupling.
Our results for the single-particle spectra
and spin/charge excitation gaps give clear indication of  
a Mott- to Peierls-insulator quantum phase transition at  
$u/\lambda\simeq1$. Quantum phonon dynamics yields pronounced effects 
in the (I)PE spectra, which might be of great importance 
for interpreting photoemission experiments 
of low-dimensional strongly correlated electron-phonon 
systems such as MX-chain compounds.\cite{TNYS90}

We gratefully acknowledge stimulating discussions 
with F. G\"ohmann, E. Jeckelmann, 
and A. P. Kampf. 
Work in Greifswald, Erlangen and Los Alamos was 
supported by Deutsche Forschungsgemeinschaft
(Focus programme SPP 1073), Bavarian Competence Network for High Performance
Computing, and US DOE, respectively. 
Numerical calculations have been performed at the HLRN
Berlin-Hannover, RRZ Erlangen and LRZ M\"unchen.


\begin{thebibliography}{[X]}
\bibitem{remark2}
In the ED calculations we separate the $Q=0$ phonon mode
(with an average occupation number of $g^2$ per site for the half-filled case)
and restrict ourselves to states with a maximum number of $M$ phonons 
for $N-1$ phonon modes. 
We achieve good convergence for  $M=22$ even in the PI regime 
but the use of leading-edge supercomputers is still mandatory
to handle the resulting Hamilton matrices with dimensions of $7 \times 10^9$.
The effect of the $Q=0$ phonon mode in the (I)PE spectra
can be easily be calculated and, of course, is included in the data shown.
\bibitem{remark3} 
All DMRG calculations were done with $m=1000$ and $n_\mathrm{b}=2$, 3
or 4 phonon pseudosites in the MI, QCP and PI regimes, respectively.
The low number of pseudosites was made possible by by performing a 
linear shift $-g\omega_0 N_{\rm el}/N$ of 
      the on-site potentials, which effectively removes the $Q=0$ phonon 
      mode. For comparison with ED data, the number
of ``real'' phonon states is $2^{n_\mathrm{b}}$.\cite{Hager}
Computational Resources ranged up to 10 GBytes of
memory and 250 CPU hours on a current Itanium 2 
machine (SGI Altix) for the 128-site PI cases.
\bibitem{remark4} 
The ground state calculation in the MI case with PBC shows a peculiar
behaviour when going beyond 32 sites. At 64 sites and larger it 
coincides exactly with the OPC ground state, showing a vanishing
bond order at site 64.  PBC are hard to
establish in our algorithm because of the particular way the DMRG
sweeps are performed. Modifications which do not show these
deficiencies are under investigation.
\end{thebibliography}
\end{document}